\documentclass[twocolumn]{aastex631}
\usepackage{amsmath}
\shorttitle{GRB Duration-Redshift Anti-Correlation}
\shortauthors{Lloyd-Ronning et al.}
\graphicspath{{./}{figures/}}

\begin{document}

\title{On the Anti-Correlation between Duration and Redshift in Gamma-ray Bursts.}

\author[0000-0002-0786-7307]{Nicole Lloyd-Ronning}
\affiliation{Los Alamos National Lab, Los Alamos, NM, USA 87545}
\affiliation{Department of Math, Enginnering, \& Science, University of New Mexico, 4000 University Dr., Los Alamos, NM, USA 87544}

\author{Jarrett Johnson}
\affiliation{Los Alamos National Lab, Los Alamos, NM, USA 87545}
 
\author[0000-0002-4854-8636]{Roseanne M. Cheng}
\affiliation{Los Alamos National Lab, Los Alamos, NM, USA 87545}

\author{Ken Luu}
\affiliation{San Francisco State University, San Francisco, CA 94132}
\affiliation{Los Alamos National Lab, Los Alamos, NM, USA 87545}

\author{Phoebe Upton Sanderbeck}
\affiliation{Los Alamos National Lab, Los Alamos, NM, USA 87545}

\author{Lailani Kenoly}
\affiliation{Los Alamos National Lab, Los Alamos, NM, USA 87545}

\author{Celia Toral}
\affiliation{Cornell University, Ithaca, NY, 14850}
\affiliation{Los Alamos National Lab, Los Alamos, NM, USA 87545}



\begin{abstract}

For gamma-ray bursts (GRBs) with durations greater than two seconds (so-called long GRBs), the {\em intrinsic} prompt gamma-ray emission appears, on average, to last longer for bursts at lower redshifts.
We explore the nature of this duration-redshift anti-correlation, describing systems and conditions in which this cosmological evolution could arise.  In particular, we explore its dependence on metallicity of a massive star progenitor, as we can securely count on the average stellar metallicity to increase with decreasing redshift. Although higher metallicity/lower redshift stars lose mass and angular momentum through line-driven winds, in some cases these stars are able to form more extended accretion disks when they collapse, potentially leading to longer duration GRBs.  We also examine how this duration-redshift trend may show up in interacting binary models composed of a massive star and compact object companion, recently suggested to be the progenitors of radio bright GRBs.  Under certain conditions, mass loss and equation of state effects from higher metallicity, lower redshift massive stars can decrease the binary separation. This can then lead to spin-up of the massive star and allow for a longer duration GRB upon the massive star's collapse. Finally, the duration-redshift trend may also be supported by a relatively larger population of small-separation binaries born in situ at low redshift.   

\end{abstract}

\keywords{}


\section{Introduction} \label{sec:intro}

There is strong evidence that most long gamma-ray bursts (GRBs), those with prompt gamma-ray durations lasting more than about two seconds, are associated with the collapse of a massive star.  Observationally, there exist definitive associations of Type Ibc supernovae with long GRBs \citep{Gal98, Hjorth03, HB12}.  Theoretically, such progenitors provide the necessary energy reservoir over the relevant timescales \citep{Woos93, MW99, WM99,WB06,WH06, KNJ08a}.\\

  However, the details of GRB progenitors are still far from resolved.  The central engine is believed to be a highly spinning black hole-disk system\footnote{There are also viable models with a neutron star as the central engine \citep[e.g,][]{Lin20}.} that launches a very relativistic jet. Internal dissipation processes in the jet produce the prompt gamma-ray emission, while the subsequent afterglow (emitted across the electromagnetic spectrum) results from the jet front's interaction with the external medium.  But which massive stars are capable of producing such powerful jets and why (i.e. what are the key parameters that allow certain massive stars to create a GRB)?  One key approach to answering this question is to examine the evolution of GRB  properties over cosmic time, and attempt to understand that evolution in the context of particular progenitor systems.  For example, the evolution of GRB jet opening angle with redshift \citep{Lu12, Las14, Las18, Las18b, LR19, Chak23} connects GRBs with massive stars in a unique way. \cite{LR20} showed that the smaller jet opening angles at higher redshifts can be quantitatively explained by a massive star's envelope collimating the jet, and is consistent with evolving IMF models reported in the literature \cite[e.g.][]{Dave08}.   \\

  In this vein, we examine the nature of the evolution of GRB prompt duration over redshift. Taking care to account for selection effects, \cite{LRAJ19} found a statistically significant anti-correlation between intrinsic duration (that is, duration corrected for cosmological time dilation) and redshift in a sample of 376 long GRBs. \cite{LR19} confirmed this in a sub-sample of radio bright GRBs with isotropic equivalent energies above $10^{52}$ ergs. Recently \cite{Chak23} further confirmed this trend, employing different techniques to account for data selection effects that could have artificially produced this intrinsic duration-redshift anti-correlation.   \\

    How might this cosmological evolution of GRB prompt duration arise?  We might, at first glance, expect that it is the {\em lower metallicity} stars (at high redshift) that would produce longer duration GRBs; higher metallicity massive stars (at lower redshifts) lose more mass and angular momentum (through line-driven winds) relative to lower metallicity stars \cite[e.g.][]{Puls08}. Under the assumption that higher angular momentum systems are capable of producing longer duration GRBs \citep{PB03,JP08, LR22}, we would then naively expect to see, on average, longer duration GRBs at higher redshifts.  That the {\em opposite} trend is seen in the data suggests this expectation warrants closer examination.   \\

In this paper, we qualitatively consider different long GRB progenitor scenarios in which this unexpected correlation could arise, focusing particularly on the role metallicity plays in determining the properties of these systems and their post-collapse GRB jets. We first examine the presence of this correlation in the context of a single star collapse.  We then consider progenitor systems with a massive star collapsing in an interacting binary system, and examine whether the evolution of binary properties with redshift can account for the observed duration-redshift anti-correlation.   Our paper is organized as follows.  In \S 2, we discuss the data and summarize the evidence for this correlation. Then, in \S 3, we describe how single star progenitors might produce this duration-redshift trend.  Recognizing the strong assumptions necessary in single star progenitor models for long GRBs, \S 4 is devoted to exploring how this trend may show up in binary progenitor models for long GRBs. In \S 5, we present discussion and summarize our conclusions.  \\

 \begin{figure*}
    \centering
 \includegraphics[width=0.62\textwidth, height=0.43\textheight]{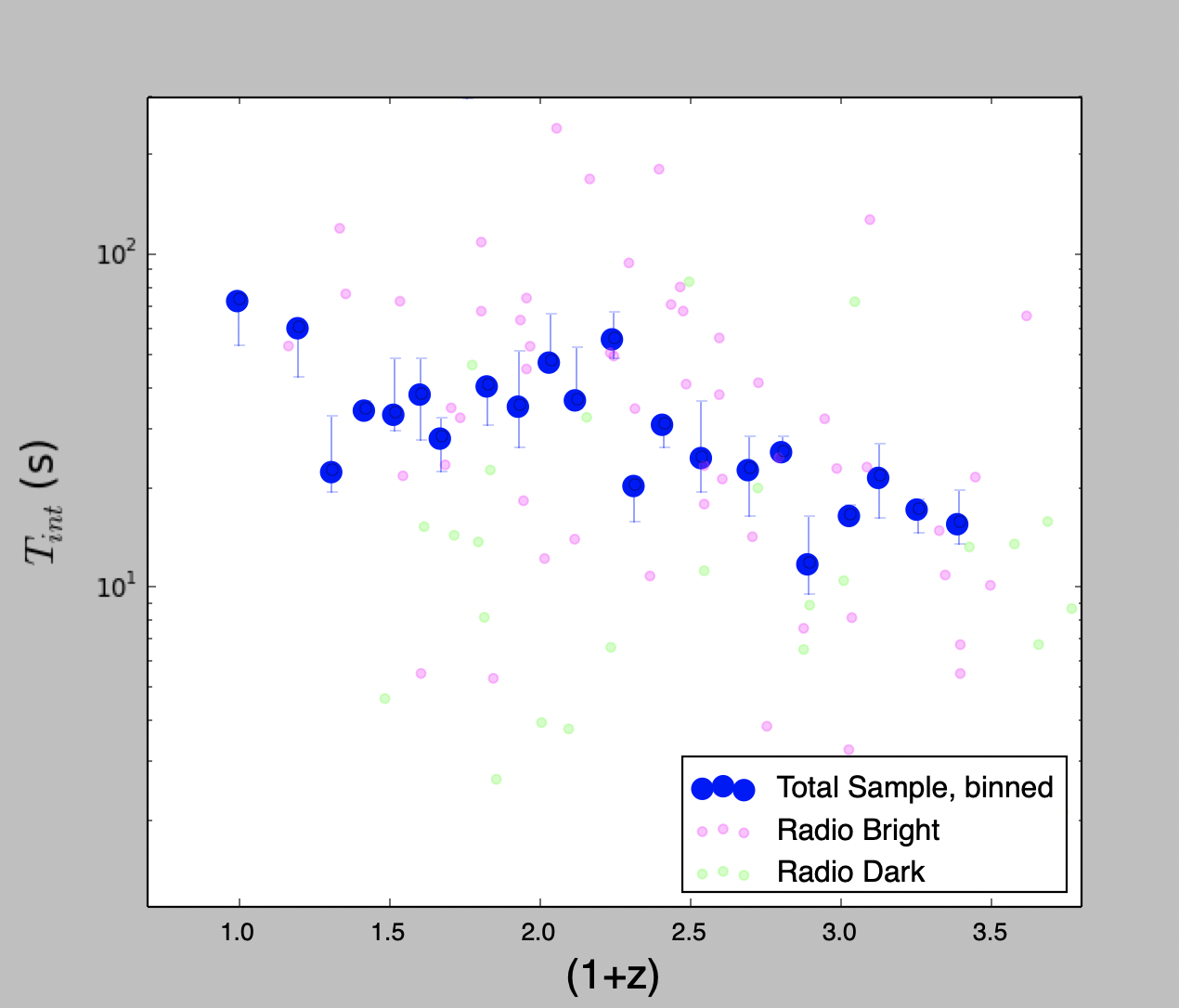} 
    \caption{Intrinsic gamma-ray duration (corrected for cosmological time dilation) as a function of redshift for 376 GRBs with redshift measurements. The data are from \cite{LR19} and binned every 20 GRBs, for clarity.  We also show the unbinned data, divided into radio bright (magenta) and radio dark (lime) subsamples.  The on average shorter prompt duration of the radio dark sample is evident.  Detailed explanations of the statistical analysis of this correlation for the whole sample as well as radio bright and dark sub-samples are found in \cite{LRAJ19,LR19,Chak23}.}
    \label{fig:tintz}
\end{figure*}

\section{Observed Correlation} 

\cite{LRAJ19} looked at a sample of 376 long GRBs with measured redshifts, and corrected the durations for cosmological time dilation to get the intrinsic duration: $T_{\rm int} = T_{90}/(1+z)$, where $T_{90}$ is the measured duration of $90 \%$ of the photon counts and $z$ is redshift. They found $T_{\rm int}$ appears on average to decrease with redshift - that is, gamma-ray bursts' intrinsic prompt durations are on average longer at lower redshifts. It is important to consider whether this correlation has a true physical origin or is a result of observational selection effects.  For example, {\em for a given luminosity}, a high redshift GRB will have a lower flux and some of the emitted flux could in principle be ``shifted" beneath the detector noise, making the GRB time profile appear shorter than it really is \citep[i.e. with an apparent smaller duration; a recent example of this effect in the context of ultra-long duration GRBs is found in][]{Huang22}. However, GRBs are {\em not} standard candles in their luminosities and in fact have much scatter in this variable; in other words, a higher redshift GRB does {\em not} necessarily imply a lower flux, and therefore this effect is substantially mitigated \citep{LBP00}. In fact, many studies have suggested GRBs are brighter at high redshifts, which would further mitigate (if not eliminate) this selection effect. Nonetheless, taking a conservative approach to account for such biases using Lyndell-Bell \citep{LB71} and Efron-Petrosian \citep{EP92} non-parametric methods, \cite{LRAJ19} found a statistically significant ($\sim 4 \sigma$,  p-value of $10^{-4}$, where the null hypothesis is no redshift evolution) anti-correlation between intrinsic duration and redshift: $T_{\rm int} \propto (1+z)^{-0.8 \pm 0.2}$.   This correlation is shown in Figure ~\ref{fig:tintz}, with data points binned (every 20 GRBs) for clarity (large blue dots).   \\

This correlation shows up even more prominently in certain sub-samples of long GRBs. \cite{LR19} found this correlation exists with even higher statistical significance in radio bright GRBs (that is, those GRBs for which radio follow-up was attempted and an afterglow was detected; note that in their analysis, they examined only those GRBs with isotropic equivalent energies above $10^{52}$ ergs). In their radio bright sub-sample consisting of 79 GRBs, they found  a $\sim 5 \sigma$ correlation (p-value $\sim 5\times 10^{-5}$), with $T_{\rm int} \propto (1+z)^{-1.4 \pm 0.3}$. Recently, \cite{Chak23}  re-analyzed this correlation with an updated, larger sample of radio bright and radio dark GRBs, using a variety of approaches to account for selection effects.  Even with fairly stringent and conservative assumptions about the biases, they found the correlation still exists, with $T_{\rm int} \propto (1 + z)^{-1.3 \pm 0.3}$ ( p-value of $10^{-6}$) for radio-bright GRB sample and $T_{\rm int} \propto (1 + z)^{-1.2 \pm 0.4}$ (p-value of $10^{-3}$) for the radio dark sample.  The unbinned radio bright (magenta) and radio dark (lime) duration-redshift data are shown in Figure ~\ref{fig:tintz}. \\


The presence of this correlation, if indeed physical, can offer important clues to the progenitor systems of long GRBs.  Because the duration of a GRB (in the context of a black hole-disk central engine) is roughly proportional to the amount of mass in the disk divided by its accretion rate\footnote{This rough proportionality is simply a dimensional analysis argument, but please see \cite{JP08} and \cite{KNJ08} for more detailed arguments.} (i.e. $T_{\rm int} \sim M/\dot{M}$), longer duration GRBs at lower redshifts imply these systems have more massive accretion disks, lower accretion rates, or both.  Of course, extrapolating from the final state of the progenitor star (or stars) to the black hole-disk system that produces the GRB jet is not straightforward \citep{Halevi2022}. However, we might expect general trends to hold - for example, a more extended massive star and/or one with an overall higher angular momentum reservoir may be able to produce a longer-lived jet upon collapse, leading to a longer duration GRB \cite[see the recent arguments in][]{LR22}. 

\section{Single Star Models}
A single massive star collapsing to a black hole-disk central engine - the so-called collapsar model \citep{Woos93, MW98, WM99} - has for some time been the canonical model for long gamma-ray bursts.  As long as the star can retain enough angular momentum to launch a jet (with enough momentum to punch through any existing stellar envelope), this progenitor provides the necessary overall energy budget to power a long GRB.  Several types of massive stars have been suggested as viable progenitors of GRBs, including red supergiants (RSG) \citep{QK12, NAK12, Wu13, Liu18}, blue supergiants (BSG) \citep{Pern18}, and Wolf-Rayet (WR) stars \citep{MM05,Georg09, Georg12}.  In particular, WR stars have been suggested as viable progenitors because of their stripped hydrogen envelopes - the stripped envelope not only makes it more viable for the jet to break free from the star but is also consistent with the lack of hydrogen lines in the spectra of supernovae associated with some GRBs\footnote{We also note the result by \cite{HH2010} who observed populations of WR stars in the host galaxies of some GRBs.}.  However, \cite{Suwa11, Nag12, Pern18} and others have shown that it is indeed possible for a jet to break out from RSG and BSG progenitors. Therefore, our arguments are applied in a fairly general way, as the details of the progenitor stars able to produce a GRB are still up for debate.\\

We consider how the duration-redshift anti-correlation might arise in isolated stellar systems, particularly as a function of metallicity (again, because this securely increases with decreasing redshift). As mentioned in the introduction, we might naively expect that, because higher metallicity stars lose more mass and angular momentum compared to lower metallicity stars \citep{WH06}, they would produce on average {\em shorter} duration GRBs, opposite to the trend we see in the data (although see \cite{Maed02} who show that anisotropic - specifically, polar enhanced - mass loss leads to an {\em increase} in stellar rotational velocity; and indeed \cite{Ol22} show observations of an 80 $M_{\odot}$ star undergoing such bipolar mass loss). \\

However, higher metallicity stars are also, on average, less dense/have larger radii than low metallicty stars \citep{MM01,Georg13}. This is established observationally in dwarf stars \citep{Hou08}\footnote{Although see \cite{Boy12} who do not find a strong dependence on metallicity for many physical properties of {\em low-mass stars}.}.  
 In the case of high mass stars, the observations are more tenuous \cite[although see][]{Harad19, Gies21}.  Though secure measurements of metallicity and radius are more difficult to come by, \cite{Far22} have shown through a number of detailed numerical simulations of massive star evolution, that lower metallicity stars are indeed more compact due to lower opacity in the stellar envelope as well as lower CNO abundances in the burning regions of the star. \cite{San17} explore the metallicity dependence of envelope inflation in massive stars and find, consistent with expectations, that, for a given mass, low metallicity stars haven  smaller radii relative to high metallicity stars (see their Figures 2 and 11).\footnote{We note the results of \cite{Chun18} whose simulations of red supergiants with 1D stellar evolution codes show that the final radii of these stars' hydrogen envelopes are largely independent of metallicity (over the range of Z=0.004 to 0.04). However, GRB models (and observations of their coincident Type Ic supernovae) generally require a stripped hydrogen envelope, and so these models may not apply to our study.}  \cite{XRM22} make an estimate of a $15 M_{\odot}$ star's radius dependence on metallicity, modelling the star with the stellar evolution code MESA, and also see a significant increase in the stellar radius with metallicity, over the range of metallicities they consider ($Z_{\odot}$ between $0.001$ and $0.4$).  \cite{Tout96, Macc04} found that the radius of a star is proportional to metallicity roughly to the $1/8$ power. \\

 What are implications of this for single star progenitors of long GRBs (and, in particular, for the lifetime of the jets they create and therefore the timescale of the prompt emission)?  We can get a handle on the duration of the GRB jet by estimating the interval over which stellar material feeds the central engine upon collapse.  In other words, the free-fall time, given by $t_{ff} \approx 4298 {\rm s} \ (R^{3}/M)^{1/2}$, can be used as a rough estimate for the lifetime of disk that will form around the central black hole.  
 As long as there is sufficient angular momentum to launch the GRB jet, simply the difference in the stellar radius at different metallicities can therefore qualitatively explain the trend we see in the data. We represent this in Figure 2, which shows the free fall time as a function of stellar radius, for stars of different masses.   Imposed on the plot are a set of stellar evolution simulations we ran with the 1D stellar evolution code MESA 
\citep{Paxton2011, Paxton2013, Paxton2015, Paxton2018, Paxton2019, paxton_bill_2021_5798242}\footnote{We use release r21.12.1}.  The models on the plot show stars of 25 solar masses (purple symbols) and 40 solar masses (lime symbols) at different metallicities ranging from $Z=0.0001$ to $Z=0.1$, where $Z$ is the mass fraction of ``heavy'' elements (heavier than helium) relative to the total mass of the gas.  Not shown on the plot are a suite of 60 solar mass models which show similar trends as we describe below.  The MESA equation of state (EOS) is a blend of the OPAL \citep{Rogers2002}, SCVH
\citep{Saumon1995}, FreeEOS \citep{Irwin2004}, HELM \citep{Timmes2000},
PC \citep{Potekhin2010}, and Skye \citep{Jermyn2021} EOSes.
Radiative opacities are primarily from OPAL \citep{Iglesias1993,
Iglesias1996}, with low-temperature data from \citet{Ferguson2005}
and the high-temperature, Compton-scattering dominated regime by
\citet{Poutanen2017}.  Electron conduction opacities are from
\citet{Cassisi2007} and \citet{Blouin2020}.
Nuclear reaction rates are from JINA REACLIB \citep{Cyburt2010}, NACRE \citep{Angulo1999} and
additional tabulated weak reaction rates \citet{Fuller1985, Oda1994,
Langanke2000}.  Screening is included via the prescription of \citet{Chugunov2007}. 
Starting with the pre-main sequence model option in MESA, we evolve the stars until the fraction of helium in the core falls below $10^{-3}$, and have done so with a range of spatial and temporal resolution parameters until convergence.
 As expected, and consistent with the results quoted in the previous paragraph, there is a clear trend of larger radius for higher metallicity.\\
 
   We note that Table 2 of \cite{WH06} also shows a general trend among all of their models of a smaller iron core mass with increasing metallicity.  This means not only is the  stellar radius more extended in higher metallicity stars but can contain more mass; there is therefore a larger reservoir of fuel to feed the black hole and sustain the jet for a longer amount of time.\\

   Metallicity has also been shown to slow accretion rate 
   \citep{WL22}. The accretion rate (due to fallback) is estimated as $\dot{M} \sim dM_{r}/dt_{ff}$, where $M_{r}$ is the mass contained in a radius $r$.  Related to our arguments above, a higher metallicity (more extended, less dense) star will have a longer fallback accretion timescale and therefore lower accretion rate \cite[see also equation 11 of][]{SI11}. Additionally, higher stellar opacity leads to larger radiation pressure that can stall accretion \citep{Fuk18}. \cite{Toy18} show how metallicity decreases the accretion rate even in the case of super-Eddington accretors\footnote{However, GRBs are known to be extremely super-Eddington accretors (more than 10 to 15 orders of magnitude above the Eddington rate), and it is not clear exactly how strong of an effect metallicity will have in this case.}.  \\



The arguments above suggest that higher metallicity isolated stars can in principle produce longer duration GRBs provided these systems have enough angular momentum to launch a jet.  This, of course, is a strong assumption (see, e.g.,  \cite{PB03} and \cite{JP08} for discussions on the difficulty of single stars retaining enough angular momentum upon collapse to produce a long duration GRB), and has led to some authors considering massive stars collapsing in an interacting binary system as a way to increase or sustain the angular momentum of the massive star upon collapse. We now discuss this correlation in the context of that scenario.

\begin{figure}
\hspace{-0.5cm}
 \includegraphics[width=0.52\textwidth]{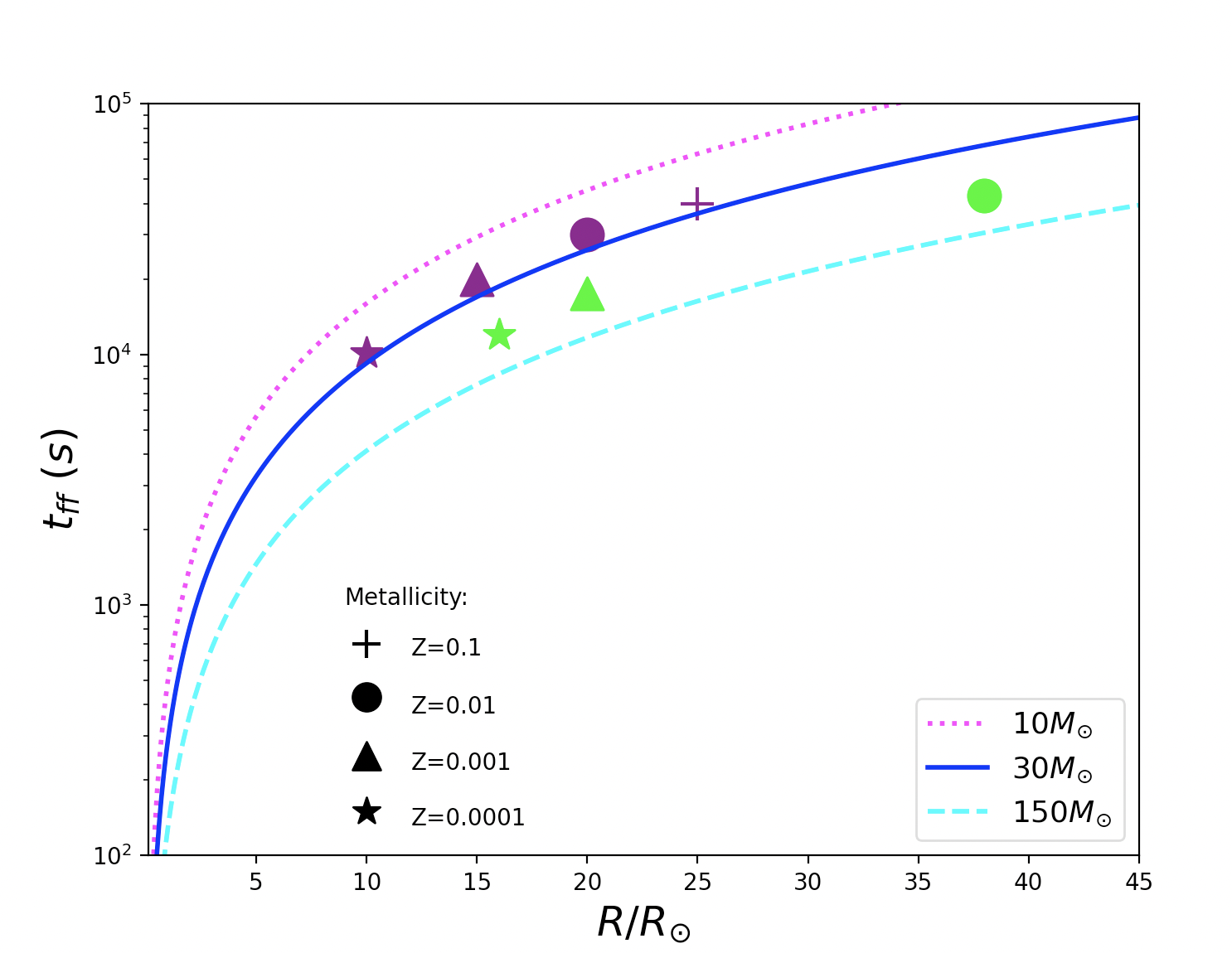} 
    \caption{Free fall time as a function of stellar radius for stars of different masses.  The plotted symbols show the radius of an evolved 25 solar mass star (purple) and a 40 solar mass star (lime) for different metallicities (denoted by different symbols indicated in the lower left of the plot), using the MESA stellar evolution code. As expected stars of larger metallicity have larger radii near the end of their evolution. We find the radius depends on metallicity roughly as: $R \propto Z^{1/8}$ for the 25 solar mass star, consistent with \cite{Tout96} and \cite{Macc04}. The relationship holds for the 40 solar mass star at lower metallicities but has a stronger dependence at higher metallicities.  We note that metallicity, as we have defined it here, is the fraction of elements heavier than helium; solar metallicity in these units is $Z \approx 0.019$}. 
    \label{fig:tff}
\end{figure}

\section{Interacting Binary Star Models}
  Recently, \cite{LR22} examined the particular case of a massive star collapsing in a tidally locked system with a black hole companion\footnote{Their work was motivated by the apparent dichotomy between radio bright and dark GRBs - as discussed above, several works have shown that the radio bright population has on average significantly longer prompt gamma-ray durations and isotropic equivalent energy compared to the radio dark population. In their model, \cite{LR22} show these systems can explain the longer duration gamma-ray emission, higher isotropic energy, and presence of radio emission in the radio bright sub-population of GRBs.}. Observationally, we know such systems exist as HMXBs.  As discussed in \cite{Tv06}, these systems (as observed in our own galaxy) have massive star companions greater than 20 solar masses and with radii of 10 to 20 solar radii. They appear to almost fill their critical Roche lobes. Their orbital periods are $< 10$ days and they have low eccentricity. 
 Motivated by this, we consider how the $T_{\rm int}-(1+z)$ anti-correlation may arise in systems that consist of a massive star in a tidally locked binary with a black hole companion. \\
 
  All of the points made in the previous section relevant to single star models will also apply to the massive star in our binary.  However, here we consider how the interaction of the massive star with its companion affects the state of the star at the end of its life.  In particular, we consider the angular momentum of the massive star right before collapse and relate this to the duration of the GRB jet.  We then examine how this depends on the metallicity of the system, in order to understand how these systems (and their GRB durations) might evolve over redshift. 

The net angular momentum (denoted by $J$) in a binary system is given by:
\begin{equation}
\begin{split}
    &J_{\rm binary}  = J_{\rm orb} + J_{*} + J_{\rm BH} = \\
   & \left[ \frac{aG(M_{*}M_{\rm BH})^{2}}{(M_{*} + M_{\rm BH})} \right]^{1/2} +  
    I_{*}\omega_{*} + I_{\rm BH}\omega_{\rm BH} 
\end{split}
\end{equation}

\noindent where $a$ is the binary separation, $G$ is the gravitational constant, $M_{*}$ is the mass of the massive star, $M_{BH}$ is the mass of the black hole, $\omega_{*}$ is the rotational rate of the massive star, $\omega_{BH}$ is the rotational rate of the black hole, $I_{*}$ is the moment of inertial of the massive star and $I_{BH}$ is the moment of inertia of the black hole.   We approximate the moments of inertia of each object as $I \approx \alpha M R^{2}$, where $\alpha$ is a constant of order unity and $R$ is the radius of the spinning object (massive star or black hole).  \\

If we assume our system is tidally locked (with zero eccentricity)\footnote{This is an arguably justifiable assumption; the tidal locking timescale is given by \citep{GLAD96}:\\
  \begin{equation}
  \begin{aligned}
        t_{lock} \approx & 10^{3} \ {\rm yr} \ (r/10^{12}cm)^{6} (\rho_{star}/10^{3}g cm^{-3}) \\
        & (\omega_{o}/10^{-5}s^{-1}) (Q/100)  (M_{BH}/10M_{\odot})^{2} (k_{2}/0.1) 
\end{aligned}
\end{equation}
    where $\rho_{star}$ is the average density of the star, $\omega_{o}$ is its initial rotational velocity, $Q$ is the so-called dissipation function of the star, and $k_{2}$ is the tidal love number. Quantities like $Q$ and $k_{2}$ are highly unknown \citep{GLAD96} and usually taken to be around $\sim 100$ and of order unity, respectively.}, the orbital frequency  $\omega_{orbit} = \omega_{*} = \omega_{BH} = \sqrt{G M_{tot}/a^{3}}$, where $M_{tot} = M_{BH} + M_{*}$.  In this special case the total angular momentum in the system can be written as:
\begin{equation}
\begin{split}
    &J_{\rm binary} =  \left[ \frac{aG(M_{*}M_{\rm BH})^{2}}{(M_{*} + M_{\rm BH})} \right]^{1/2} + \\
    & \left[ \frac{G(M_{*} + M_{\rm BH})}{a^{3}} \right]^{1/2} \left[ \alpha_{\rm BH}M_{\rm BH}R_{\rm BH}^{2} +  \alpha_{*}M_{*}R_{*}^{2} \right] 
    \end{split}
\end{equation}

  Again, here our goal is to consider how the angular momentum of the massive star (inextricably linked to the intrinsic duration of the prompt GRB) depends on the parameters of this interacting binary system - particularly metallicity - and how this could lead to the evolution of GRB prompt duration as a function of redshift.   \\
  
  Just as in the single star case, a higher metallicity massive star in such a system will lose more mass due to line driven winds, compared to one of low metallicity. This  non-conservative mass loss can lead to the binary widening \citep{Floor22}, and decrease the binary interaction (lessening the potential spin-up of the massive star)\footnote{although we note again that mass loss could be anisotropic as discussed in \cite{Maed02}, which can lead to spin up of the massive star.  As mentioned in the previous section, such anisotropic mass loss is observed in a massive star in an interacting binary system closely related to the ones we consider here \citep{Ol22}.}.  Therefore, as in the single star case, we might at first glance expect higher metallicity systems to have lower angular momentum and therefore shorter duration GRBs.  However, this ignores a number of other subtle factors that can decrease the binary separation, and lead to spin-up of the massive star.  For example, a higher metallicity massive star in a binary system may also be more likely to undergo conservative mass loss through Roche lobe overflow, while these stars are less compact, and this can serve to tighten our binary.  Drag forces by ambient gas can also play an important role and lead to a decrease in the binary separation.  We now discuss scenarios in which (ultimately metallicity-dependent) changes in mass and radius affect the binary system and play a role in the duration of a GRB. 
  
   In short, higher metallicity serves to increase both conservative and non-conservative mass loss (the former of which will lessen the binary separation, the latter which will increase it), and there is a competition at play between these two effects.  Additionally, higher metallicity systems can play a complicated role in other effects such as anisotropic mass loss and the role of drag forces (both of which decrease our binary separation). Below we delve into this issue in more detail as we attempt to understand the role of metallicity in the evolution of binary separation. \\

 \subsection{Evolution of Binary Separation}
  How the separation between two stars in a binary system evolves over the lifetime of each star is a complicated problem with many possible outcomes, depending on the physical properties the stars, the initial binary separation, the nature of each star's mass loss, and the system's environment. \\
  
  It is crucial to point out that we observe a number of HMXB systems (similar to the systems we consider here) which are indeed undergoing a decrease in their orbital separation - for example, Cen X-3 \citep{Kelley83}, SMC X-1 \citep{Lev93}, LMC X-4 \citep{Lev00}, and seven additional HMXBs \citep{Fal15}, all with orbital decay timescales on the order of about a half a million years \cite[see Table 3 of][]{Fal15}. \\


 For the idealized case of two point-like masses, we can write the normalized change in binary separation as \citep{PY14}:\\
  \begin{equation}
     \begin{aligned}
     \frac{\dot{a}}{a} & = \\
     & -2\frac{\dot{M_{*}}}{M_{*}}\left[1+(\alpha-1)\frac{M_{*}}{M_{BH}} - \frac{\alpha M_{*}}{2 M_{tot}} \right] 
      + \ 2\frac{\dot{J}_{orb}}{J_{orb}}
     \end{aligned}
 \end{equation}
 \noindent where $\dot{M}_{*} < 0$ (i.e. the star is losing mass).  Note that $\dot{M}_{*}$ and $\dot{J}_{orb}$ are not independent variables.  The parameter $\alpha$ is the fraction of the lost mass from $M_{*}$ that {\em leaves} the system.  That is, for $\alpha = 0$, all the mass lost from our star stays in the binary system; for $\alpha = 1$, all of it leaves the binary system.  \\
  
 In the case of conservative mass transfer ($\alpha = 0$ and $\dot{J}_{orb} = 0$), when all of the massive star's lost mass is deposited onto the black hole, the ratio of initial to final binary separation is given by \citep{PY14}:
  
  \begin{equation}
      \frac{a_{f}}{a_{i}} = \left(\frac{M_{BH,i}M_{*,i}}{M_{BH,f}M_{*,f}}      \right)^{2}
  \end{equation}\\
  
  \noindent where $i$ and $f$ denote initial and final states, respectively. For a massive star that is more massive than the black hole companion, as we reasonably assume is the case here \citep{2003ApJ...591..288H,2022arXiv221112153R}\footnote{We note that if the binary system was born in situ (rather than through dynamical capture), the black hole companion must be the result of the more massive star of the two original stars in the binary.  However, we assume the death of the first massive star left a remnant black hole that is less massive than the second massive star (the one we consider here as the GRB progenitor).},  the binary orbit will shrink as $M_{*}$ transfers mass to the black hole. As a result, the system's angular frequency $\omega$ increases, and (because the system is tidally locked) the massive star's rotation rate increases.  As mentioned above, this provides a larger angular momentum reservoir which can lead to a longer duration GRB \citep{LR22}.  In such a scenario, more mass loss would therefore lead to a longer duration GRB.   However, conservative mass loss applies primarily in the the case of Roche lobe overfill. In general, mass loss from winds (either isotropic mass loss or pole-dominated anisotropic mass loss)is {\em not} conservative and some mass will escape the system. \\


  
  In the case when all of the mass directly leaves the system (perfect non-conservative mass loss with $\alpha = 1$ or the so-called Jeans mode), the binary separation will increase in order to conserve angular momentum. However, there are situations of non-conservative mass loss that lead to reduced orbital separation. For example, in the case when accretion from the massive star onto the black hole exceeds the Eddington limit, matter can be expelled from the system by radiation pressure.  In this case, angular momentum is depleted from the accretor and the orbital separation can exponentially shrink with mass loss \cite[see equation 43 of][]{PY14}: 
  
  \begin{equation}
      \frac{a_{f}}{a_{i}} = \left(\frac{M_{BH,i}}{M_{BH,f}}\right) \left(\frac{M_{*,i}}{M_{*,f}}\right) {\rm exp} \left( -2 \frac{M_{*,i} - M_{*,f}}{M_{BH}} \right)
  \end{equation}
 
\noindent Hence, in a realistic system there are a number competing effects when we consider the role of metallicity driven mass loss as it affects our binary system and ultimately the spin of the massive star.  \\

 Finally, we note that drag forces can also have an important effect on the system's evolution.
 \cite{Schr21} showed that, under the assumption of non-spinning stars, line driven winds can either tighten or widen the binary depending on the mass ratio of the binary, as well as the ratio of the wind velocity to orbital velocity.  They found that for slower winds (relative to the orbital velocity) and higher mass companions, drag forces serve to shrink the binary separation (see their Figure 1). Interestingly, \cite{Fal15} report inferred wind speeds for the massive stars in 10 HMXBs on the order of $1000 km/s$ (see their Table 9).  Meanwhile, the orbital velocities for our closely separated, tightly bound binaries are about an order of magnitude higher than this, satisfying the necessary velocity ratio condition that leads to binary separation decrease.\footnote{In our systems, we are typically considering a massive star with a compact object companion whose mass is expected to be less than that of the massive star and so - although drag forces may play an important role in the binary separation evolution - the \cite{Schr21} calculations may not apply to our models.} \\

 \subsubsection{The Effect of Stellar Expansion}
  
  As mentioned in the previous section on single star progenitors, higher metallicity stars are more extended in radius. A more extended star can have a higher moment of inertia and therefore - for a given rotation rate - higher angular momentum. Therefore, as the star expands, the binary separation will decrease (in order to conserve angular momentum in the system as a whole).   \cite{Lev93,Lev00} derive an expression for this change in orbital period $P_{orb}$ as a function of the change in the moment of inertia of an expanding star:
\begin{equation}
    \frac{\dot{P}}{P} = \frac{3\dot{a}}{2a} \approx \frac{\omega_{BH}d{\rm ln}I_{*}/dt}{\omega(\mu a^{2}/3I_{*} - 1)}
\end{equation}
 
\noindent where  $\mu$ is the binary reduced mass\footnote{Recall, in our scenario, $\omega_{BH} \sim \omega$ (the orbital angular frequency), which simplifies the equation.}
This equation only accounts for radial expansion and assumes the mass stays constant.  As mass is lost, the moment of inertia will decrease accordingly. 
 However, if the normalized mass lost to winds is less than twice the normalized change in radius due to stellar expansion, 
\begin{equation}
 \Delta M/M < 2 \Delta R/R 
 \label{eq:delM}
\end{equation} 
\noindent then the star's expansion will ``win'', and serve to decrease the separation in our binary system. Again, in our tidally locked model, this leads to spin-up of our massive star.  

The rate of mass loss due to line-driven winds is a complicated and unresolved problem.  However, it is reasonable to expect the condition in equation~\ref{eq:delM} holds, while stellar winds are often weaker than simple models predict. We refer the reader to  the review by \cite{Smith14} for an in-depth look at this issue, and also \cite{Full06} who showed, observationally, that stellar models tend to overestimate mass loss due to winds by a factor of 10 or more, when realistic conditions (such as clumping) are not considered.  \\

 \begin{figure}
 
 \hspace{-1cm} \includegraphics[width=10cm, height=13.5cm]{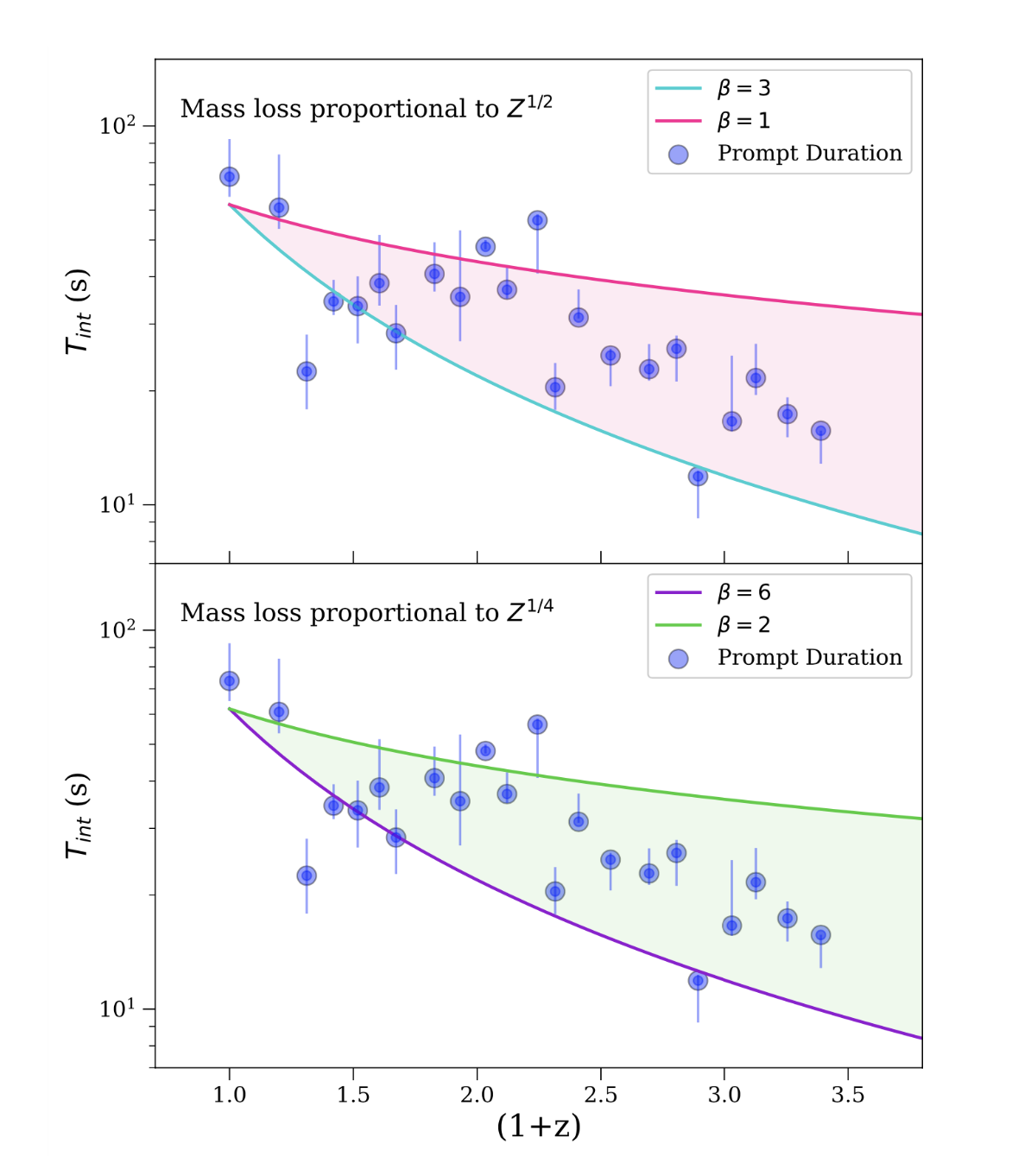} 
 \caption{{\bf Upper Panel:} Range of $\beta$ values that accommodate the trend seen in the data, in the context of a massive star collapsing in an interacting binary. The parameter $\beta$ describes how the binary orbit shrinks as a function of mass loss, where we assume mass loss depends on metallicity to the 1/2 power. {\bf Lower Panel:} Same as the upper panel, but here we assume mass loss depends on metallicity to the 1/4 power.}
    \label{fig:tmodel}
    
 \end{figure}

In short, there is a complicated interplay of many different factors related to changes in mass and stellar radius that can serve to either increase or decrease the binary separation. In the next section, we focus on the consequences of the cases in which binary separation decreases with (metallicity-driven) mass loss and relate this to the observed anti-correlation between GRB duration and redshift.  
  
\subsection{Reproducing the Anti-correlation}
As discussed above, there exist various scenarios in which the binary separation will decrease for systems consisting of a massive star with a compact object companion.  In our model, where the tidal locking timescale is relatively short compared to other timescales in the system (e.g. orbital decay time, lifetime of star), the decreased separation will lead to a more rapidly spinning massive star, which is capable of producing a longer duration gamma-ray burst. We would like to understand how the increase in average metallicity over cosmic time may affect properties of interacting binary systems, and how this might lead to the observed anti-correlation between prompt GRB and redshift.  \\

 We parameterize the binary system's separation decrease as function of donor star's mass loss as follows:  
\begin{equation}
    \frac{a_{f}}{a_{i}} \propto (\frac{\Delta M_{*}}{M_{*}})^{-\beta}
    \label{eq:afai}
\end{equation}
\noindent where $\beta > 0$ parameterizes the extent to which the orbit shrinks as a function of mass loss. Once again, when considering how equation~\ref{eq:afai} describes separation as a function of cosmic time, we will consider the variables' dependencies on metallicity. 

As discussed in the previous section, mass loss from both winds and Roche lobe overflow will depend on metallicity.  Several groups have parameterized the former relationship, showing that mass loss from winds is $\Delta M \propto Z^{1/2}$ \citep{Vink01, YLN06, WH06}. Mass loss due to RLOF is not as well parameterized and depends on a number of factors in the system; however, given that the radius of the star depends on metallicity roughly to the $1/8$ power as discussed above \citep{Tout96,Macc04}, then the mass loss dependence on metallicity due to RLOF can be estimated as $\Delta M \propto Z^{2/8}$ to $Z^{3/8}$ (the former considers only surface mass loss). Substituting these relationships into equation~\ref{eq:afai}, we find then that $\frac{a_{f}}{a_{i}} \propto Z^{-\beta/2}$ to $Z^{-\beta/4}$.

Because the angular momentum of the massive star is inversely proportional to binary separation under the tidal locking assumptions of our model, $J_{*} \propto a_{f}^{-3/2}$, we find a relationship between the star's angular momentum and its metallicity (in this special case of an interacting binary):
\begin{equation}
   J_{*} \propto Z^{3\beta /4}
   \label{eq:Jstar}
\end{equation} 


Equation~\ref{eq:Jstar} is considering the angular momentum of the star as it depends on binary separation in a tidally locked system. At the same time, the star's winds will carry away angular momentum and we need to consider the importance of that depletion in these arguments. Assuming that the mass lost from winds is much less than the total mass in the system \citep{Full06, Smith14}, we can estimate the change in angular momentum due to winds as $\Delta J_{wind \ loss} \approx \Delta M_{*} R_{*}^{2} \omega_{*}$, and that gained from spin up as $\Delta J_{spin \ up} \approx M_{*} R_{*}^{2} \omega_{*} (\Delta(a)/a)$. The net gain in angular momentum of the star (due to orbital separation decrease) is then $\Delta J_{*} = J_{spin \ up} - J_{wind \ loss}$.  The assumption that angular momentum loss due to winds is negligible compared to the gain due to decreasing binary separation and spin-up of the star, amounts to the following condition:


\begin{equation}
    \frac{\Delta J_{\rm wind \ loss}}{\Delta J_{\rm spin \ up}} \approx (\frac{\Delta M}{M})(\frac{a}{\Delta a }) < 1
\end{equation} 

\noindent or

\begin{equation}
    \frac{\Delta a }{a} > \frac{\Delta M}{M} 
    \label{eq:dela}
\end{equation} 

\cite{Fal15} provide estimates of mass loss and observed orbital period change in the types of binary systems we are considering.  From the data in their Tables 1, 3, and 9, we see that the condition in equation~\ref{eq:dela} above generally holds - in other words, the fractional change in binary separation is larger than the fractional mass loss in these systems and we can invoke equation~\ref{eq:Jstar} going forward.  Then, using $T_{dur} \propto J_{*}^{7/4}$ \citep{LR22} and equation~\ref{eq:Jstar}, we find:
\begin{equation}
  T_{\rm int} \propto Z^{21 \beta/ 16}  
  \label{eq:TintMet}
\end{equation} 

Now consider $Z$'s dependence on redshift. Figure 2 of \cite{Fyn06} and Figure 14 of \cite{Mad14} show how cosmic metallicity increases with decreasing redshift.  Although there is ample scatter in the data, we can approximate the relationship roughly as $Z \propto (1+z)^{-0.4}$. Substituting that relation into equation~\ref{eq:TintMet} leads to:
\begin{equation}
    T_{\rm int} \propto (1+z)^{-21 \beta /40} \approx (1+z)^{-0.5 \beta}
    \label{eq:Tintz}
\end{equation}
\noindent assuming mass loss is roughly proportional to $Z^{1/2}$.  For mass loss proportional to $Z^{1/4}$, we find:
\begin{equation}
    T_{\rm int} \propto (1+z)^{-21 \beta /80} \approx (1+z)^{-0.25 \beta}
\end{equation}

  As discussed in section 2, the observed anti-correlation between duration and redshift for the larger GRB sample (that is, not separating into radio bright and dark sub-samples) is approximately parameterized as $T_{dur} \propto (1+z)^{-0.8 \pm 0.25}$. We conservatively choose constraints on $\beta$ within the $2\sigma$ error bars and find this implies $1 \lesssim \beta \lesssim 3 $ for $\Delta M \propto Z^{1/2}$ and $2 \lesssim \beta \lesssim 6 $ for $\Delta M \propto Z^{1/4}$. These two metallicity dependencies and our model parameter spaces are shown in the two panels of Figure~\ref{fig:tmodel}.  \\

\section{Discussion and Conclusions}
In this paper we have examined scenarios in which the average intrinsic prompt duration of GRBs increases with decreasing redshift, as the data suggest.  As metallicity is a variable we can securely count on increasing over cosmic time \citep{Lang22}, we have in particular linked this cosmological evolutionary trend to the metallicity of the system. We have considered both single massive stars and massive stars collapsing in interacting binary systems as our GRB progenitors. \\

Our main results are summarized as follows:\\
\begin{itemize}
    \item {\bf Single massive star} GRB progenitors may produce the duration-redshift anti-correlation as a result of higher metallicity stars  having larger radii and being able to form more extended accretion disks with lower accretion rates. We present 1D stellar evolution simulations that support this picture. Provided there is enough angular momentum to launch a jet, the more extended  stellar radius of a higher metallicity star can ultimately lead to longer duration GRB jets at lower redshift.
    \item {\bf Massive stars in interacting binary systems} may produce the duration-redshift anti-correlation through metallicity-dependent mass loss scenarios (as well as metallicity-dependent equations of state) that lead to a decrease in the binary separation for higher metallicity, low-redshift systems. This can produce a more pronounced spin-up of the massive star and therefore longer duration GRBs at lower redshifts. Our key assumption, when connecting the binary separation to the angular momentum of the massive star, is that the tidal locking timescale is shorter than other timescales in our system's evolution, so that the binary separation sets the rotation rate of the massive star ($\omega_{*} = \sqrt{G M_{tot}/a^{3}}$). 
\end{itemize}

Both our single star and binary star pictures have made a number of simplifying assumptions.   In our single star models, we have given relatively general arguments and not focused on details of particular progenitors such as RSGs and RBGs vs WR stars.  We do note that \cite{Crow07} argues that WR stars are more common at high metallicity and so further examining the correlation in the context of details of different types of massive stars could further illuminate the nature of this correlation.  We have also made the assumption that the single/isolated star has enough angular momentum to launch a GRB jet. In our binary scenario, we've implicitly assumed a larger angular momentum resevoir present in the massive star at the end of its life will lead to a more rapidly spinning black hole-disk system (produced upon the star's collapse), and therefore a longer GRB.  There are many complicated steps connecting the spin of the massive star at the end of its life to the angular momentum of the black hole-disk system after collapse \citep{Qin18}.  Going forward we plan to examine our models through more detailed simulations of these systems, particularly the end-of-life state of massive stars in interacting binaries. We will be able to test the range of values that are valid for our parameter $\beta$, while also including effects of the stellar equation of state and the star's expansion. We would also like to further examine the dichotomy between radio bright and radio dark GRBs, particularly the conjecture that GRBs with radio afterglows, which have longer prompt gamma-ray duration, are the result of massive stellar collapse in such interacting binary systems \citep{LR22}.  As the data improve and we get a better handle on how to divide the radio bright and dark sub samples, this and similar correlations may be key to unlocking any fundamental differences in the underlying progenitor systems of GRBs.  \\

We note again that we are only considering black hole-disk central engines, and not magnetar-driven engines for our GRBs.   In GRBs with magnetar central engines, the GRB can be initially powered by accretion onto a newly born neutron star (e.g. \cite{Bern13, Bern15}; extended emission (like a plateau phase) may result from the spin down power of the magnetar \citep{ZM01}.  It is not immediately clear whether these systems would be able to produce the anti-correlation between duration and redshift, although many of the arguments we make in this paper may apply to these systems as well. \\

 Finally, evolution of the population statistics of interacting binaries may also explain the duration-redshift anti-correlation.  If there exists a larger population of interacting binary systems (of the type we discuss in \S 4 above) at lower redshifts relative to high redshifts, this could account for the trend we see in the data.  Indeed, \cite{LMB21} predict there are fewer close binaries at high redshift, although they examine only PopIII stars in their study.  On the other hand, \cite{Moe19} and \cite{Bate19} suggest the close binary fraction is anti-correlated with metallicity.  The latter in particular explore the dependence of fragmentation on metallicity and find - for low mass stars - there is a {\em higher} close binary fraction at {\em lower} metallicity (although \cite{Kur22}, who explore binary separation in situ, suggest disc fragmentation is not a dominant pathway for binaries of close separation).   We emphasize that it is not just the interacting binary fraction but the {\em overall} binary fraction that may depend on metallicity.  For example, \cite{Neug21} compared the red supergiant binary fraction in M31 and M33, and showed that it indeed depends on metallicity, with higher metallicity environments leading to a higher red supergiant binary fraction (while, again \cite{Moe19}, for example, showed the binary fraction for solar-type stars {\em decreases} with metallicity. Needless to say, this is a complex relation that warrants further examination.  A closer look at the viability of this pathway to explain the duration-redshift trend we see in the data requires detailed and informed population synthesis modelling, which we explore in a future paper.

 \section{Acknowledgements}
 We thank the anonymous referee for many useful comments that improved this manuscript.  We thank Anne Noer Kolborg and Krystal Ruiz-Rocha for helpful conversations related to cosmic metallicity evolution, as well as Angana Chakrobarty and Maria Dainotti for useful conversations related to the anti-correlation between duration and redshift.  We thank Gabriel Casabona and Shane Larson for insightful conversations on population synthesis.  We thank the MESA community and developers for their help and the many excellent resources in support of this code.  
This work was supported by the US Department of Energy through the Los Alamos National Laboratory.  Los Alamos National Laboratory is operated by Triad National Security, LLC, for the National Nuclear Security Administration of U.S. Department of Energy (Contract No. 89233218CNA000001).   Research presented in this article was supported by the Laboratory Directed Research and Development program of Los Alamos National Laboratory under project number 20230115ER. 
We acknowledge LANL Institutional Computing HPC Resources under project w23extremex.  
This research used resources provided by the Los Alamos National Laboratory Institutional Computing Program.
LA-UR-22-32979


\bibliography{refs} 
\bibliographystyle{aasjournal}



\end{document}